\documentclass[a4paper, amsfonts, amssymb, amsmath, reprint, showkeys, nofootinbib, twoside,longbibliography, aps]{revtex4-1}
\usepackage[english]{babel}
\usepackage[utf8]{inputenc}
\usepackage[detect-all]{siunitx}
\usepackage{comment}
\usepackage[normalem]{ulem} 
\usepackage[colorinlistoftodos, color=green!40, prependcaption]{todonotes}
\usepackage{amsthm}
\usepackage{mathtools}
\usepackage{physics}
\usepackage{xcolor}
\usepackage{graphicx}
\usepackage[left=23mm,right=13mm,top=35mm,columnsep=15pt]{geometry} 
\usepackage{adjustbox}
\usepackage{placeins}
\usepackage[T1]{fontenc}
\usepackage{lipsum}
\usepackage{csquotes}
\usepackage[pdftex, pdftitle={Article}, pdfauthor={Author}]{hyperref} 

\usepackage{soul,xcolor}

\begin{document}


\title{Optical losses in pure crystalline silicon in the IR band measured using WGM microresonators}

\author{Artem E. Shitikov\textsuperscript{1}}
\email{a.shitikov@rqc.ru} 
\author{Tatiana S. Tebeneva\textsuperscript{1}}
\author{Oleg V. Benderov\textsuperscript{2,3}} 
\author{Dmitry~A.~Mylnikov\textsuperscript{2}}
\author{Valery E. Lobanov\textsuperscript{1}}
\author{Dmitry A. Chermoshentsev\textsuperscript{1}}
\author{Igor A. Bilenko\textsuperscript{1,4}} 

\affiliation{\textsuperscript{1}Russian Quantum Center, 143026 Skolkovo, Russia}
\affiliation{\textsuperscript{2}Moscow Institute of Physics and Technology, 141701, Dolgoprudny, Russia}
\affiliation{\textsuperscript{3} Institute for Theoretical and Applied Electromagnetics, Russian Academy of Science, 125412, Moscow, Russia}
\affiliation{\textsuperscript{4}Faculty of Physics, Lomonosov Moscow State University, 119991 Moscow, Russia}

\begin{abstract}
The need for semiconductor technology for crystalline silicon of the highest purity and homogeneity has provided samples exhibiting low optical absorption in the infrared range. Such silicon has become the basis for photonic elements in the telecommunication band, including high-Q microresonators, which are particularly important. However, at longer wavelengths, the loss mechanisms have not yet been sufficiently studied. At the same time, this range is extremely important, especially for biological and medical applications and for fundamental research. We used optical microresonators with whispering gallery modes made from various types of silicon crystals as a tool to study the loss mechanisms. 
The study involved the pump wavelengths 1.5, 2.6, 6.1, and 8.6 \textmu m and the maximum measured Q-factors were $1.5\cdot10^9$, $5\cdot10^8$, $1.6\cdot10^7$, and $5\cdot10^4$, respectively. We showed that the conductivity type does not noticeably influence the optical losses, while resistivity and the growing method are defining factors. Our study confirms the utility of WGM microresonators as loss measurement tools and provides significant potential for the development of silicon microresonator-based photonics in the mid-IR band.
\end{abstract}

\maketitle

\section{Introduction}

Whispering gallery mode (WGM) microresonators are of particular interest in both fundamental research and cutting-edge technologies. The combination of a high-Q-factor and small mode volume is a key factor determining their wide applicability \cite{braginsky1989quality}. Their applications include laser stabilization \cite{liang2015ultralow, kondratiev_recent_2023}, Kerr comb generation \cite{herr2014temporal}, microwave generators \cite{IlchenkoCombRadio}, spectroscopy \cite{Farca:07, Rasoloniaina:15,zhu_optothermal_2011}, biological and chemical sensing \cite{Zhu2010, Ward:11}. 

The WGM microresonators exhibit substantial potential for mid-IR applications. This bandwidth covers molecular fingerprints along with the atmospheric transparency window at 3–5 \textmu m.
The Q-factor exceeding $10^8$ was demonstrated in fluorides (BaF$_2$, CaF$_2$, SrF$_2$) at 4.5 \textmu m. However, the same work demonstrated that multi-phonon absorption limits Q at a 10 million value for MgF$_2$ \cite{lecaplain2016mid}. Using a high-Q CaF$_2$ microresonator at 4.5 \textmu m, self-injection locking was applied to demonstrate the stabilization of a QCL laser \cite{doi:10.1002/lpor.201500214}. 
Kerr frequency combs in high-Q MgF$_2$ and CaF$_2$ microresonators with a Q-factor exceeding $10^8$ at 4.5 \textmu m were demonstrated in \cite{Savchenkov:15}. Thus, the Q-factor of fluorides is limited for the wavelength longer than 4.5 \textmu m due to multiphonon absorption, and at the longer wavelength the limitations are stronger \cite{grudinin2016properties}.
Alternative perspective materials for mid-IR are chalcogenide and fluoride glasses because of their wide transparency window in the infrared band. The WGM microresonators made of these glasses are widely applied for lasing, sensing applications, nonlinear conversion \cite{ZBLAN_applic_1, Way:12, Vanier:13}. The highest Q-factor measured in ZBLAN microspheres at 2.64 \textmu m exceeds $10^8$ \cite{Tebeneva:22}. 

Semiconductor materials, as silicon and germanium, are among the most attractive ones for the development of WGM microresonators in the mid-IR range. Both materials have wide transparency windows, large refractive indices, and large nonlinear optical coefficients ($n_2$ = $6\cdot10^{-17}$\,m$^2$/W for germanium \cite{Zhang_nonlinear, Li:80, Kaplunov:19} and $n_2$ = $1\cdot10^{-17}$ m$^2$/W for silicon \cite{Hurlbut:07}). WGM microresonators made of crystalline silicon have demonstrated ultra-high-Q at the telecom wavelength \cite{Shitikov:18}, in THz \cite{vogt2020terahertz} and mid-IR \cite{shitikov2020microresonator}. To date, the frequency comb in a silicon on-chip microresonator was shown at 3.5 \textmu m. However, the comb was generated in the microring with a relatively modest Q-factor $\sim 10^5$ \cite{yu2016mode}. 
Their CMOS compatibility further enhances the attractiveness of both materials for fabrication of integrated mid-IR photonic devices \cite{yu2016mode, 8010811, 668668}. 

Silicon is the major material for microelectronics. So that, high-purity samples are widely available. In particular, future generations of gravitational wave detectors are likely to use a wavelength of 2.06 \textmu m and silicon test masses \cite{shapiro2017cryogenicalLIGO, adhikari2019astrophysical}. As a semiconductor material, silicon experiences strong two-photon absorption up to 2.3 \textmu m; at longer wavelengths weaker multi-photon absorption occurs \cite{wang2013multi}. At wavelengths longer than 8 \textmu m multi-phonon absorption becomes a dominant loss mechanism \cite{hass1977residual}. Therefore, to realize the potential of silicon microresonators, it is reasonable to focus on the 2-8 \textmu m bandwidth. 

This study is dedicated to investigating optical losses in crystalline silicon using WGM microresonators. We measured Q-factors of microresonators made of crystalline silicon of different types (n-doped, p-doped, different resistivity, different crystal growth methods) at wavelengths of 1.5, 2.6, 6.1 and 8.6 \textmu m.  All microresonators are made using the same fabrication and surface polishing techniques. We used silicon with high and low resistivity to extract the Rayleigh scattering influence. Our measurements show that the resistivity is a key parameter for obtaining high Q-factors. Samples with high resistivity (6 $\rm{k\Omega \cdot m}$) exhibited the Q-factor exceeding $10^9$ at 1.5 \textmu m. At 6.1 \textmu m the Q-factor decreased to $1.4\cdot10^7$. However, to the best of our knowledge, this is the highest reported value measured at such long wavelength. For samples with 100 $\rm{\Omega \cdot m}$ resistivity, the values of the Q-factor did not exceed $2\cdot10^6$ at both 1.5 and 6.1 \textmu m. Furthermore, the samples grown by the high-floating zone technique exhibit higher Q-factors than those grown by the magnetic Czochralsky technique with the same resistivity. 

\section{Optical losses in silicon microresonators}
The quality factor $Q$ for a microresonator mode can be written in terms of the loss coefficient $\alpha$:
\begin{equation}
Q = \frac{2\pi n}{\alpha \lambda},
\label{eq:Q}
\end{equation}
where $n$ is the refractive index, $\lambda$ is the wavelength. The optical losses, in turn, can be represented as a sum: 
\begin{equation}
\alpha = \alpha_{r} + \alpha_{b.a.} + \alpha_{s.a.} + \alpha_{b.s.} + \alpha_{s.s.}  + \alpha_c,
\label{eq:alpha}
\end{equation}
where $\alpha_{r}$ is the radiation losses on the curved surface, $\alpha_{b.a.}$ and $\alpha_{s.a.}$ are the absorption in the material and on the surface, $\alpha_{b.s.}$ and $\alpha_{s.s.}$ are the scattering inside the material and on the surface, $\alpha_c$ is the coupling losses.  
$\alpha_{r}$ plays a significant role when a microresonator circumference of several tens of wavelengths. Scattering dominates when there are inhomogeneities in the material or some imperfections on the surface, such as dust particles or roughness on the surface. Resonant backscattering can excite the counter-propagating wave. If the energy exchange rate between the two waves exceeds the relaxation time, resonance splitting can be observed.
The scattering losses are expected to decrease with the increasing wavelength. Rayleigh scattering at the surface has been analyzed both theoretically and experimentally \cite{gorodetsky2000rayleigh, vernooy1998high, lin2018dependence}. In \cite{gorodetsky2000rayleigh} for the TE mode, the surface scattering is given by:
\begin{equation}
\alpha_{s.s.} = \frac{16\pi^3 n^2 \sigma^2 B^2}{3 a \lambda^4},
\label{eq:alphass} 
\end{equation}
where $\sigma$ is the variance of the surface roughness and $B$ is the correlation length of the surface roughness. It can be seen that the optical losses decrease as $\lambda^4$.

Absorption in the bulk or on the surface leads to heating of the pumped resonator, which shifts its eigenfrequencies. Bulk absorption, caused by the interband transitions of charge carriers and transfer of energy to the vibrational modes of the crystal lattice, can be greatly increased by impurities and lattice defects. Linear bulk losses in silicon can be evaluated from resistivity ($\rho$) \cite{degallaix2013bulk_si_absorp} using equation $\alpha_{\rho} = (0.0454 [\Omega])/\rho$. In this research, we used several types of silicon with different resistivity values to compare and fully assess the impact of different mechanisms of losses on the Q-factor at different wavelengths. In pure materials with low defect density fundamental optical absorption is caused by different mechanisms for short and long wavelengths. In the long-wave region, absorption is determined by the lattice vibrational spectrum and anharmonicity through multiphonon absorption. In the short-wave region, absorption is determined by the ultraviolet absorption edge (the Urbach tail). 
 
 In the case of a semiconductor material with the bandgap value comparable with a photon energy, a nonlinear power-dependent absorption may be significant, including the two-photon absorption ($\alpha_{TPA}$) \cite{bristow2007two_photon_abs, Tebeneva:24} and the free-carrier absorption ($\alpha_{FCA}$) \cite{Claps:04}: $\alpha_{b.a.} = \alpha_{l.b.a.} + \alpha_{TPA} + \alpha_{FCA}$.
In silicon, these processes can contribute significantly at wavelengths shorter than 2.3 \textmu m.
At wavelengths longer than 2.3 \textmu m, three-photon absorption becomes relevant instead of TPA, and four-photon absorption becomes dominant at wavelengths longer than 3.6 \textmu m \cite{wang2013multi}. %
The water layer on the surface of the microresonator can lead to additional surface absorption. The optical losses in this layer can be estimated as \cite{vernooy1998high}:
\begin{equation}
\alpha_{s.a.} = 4\sqrt\frac{\pi}{a \lambda} n^{5/2} \beta_\omega (\lambda) \delta ,
\label{eq:alphawa} 
\end{equation}
where $a$ is the radius of the microresonator, $\beta_{\omega}(\lambda)$ is the wavelength-dependent absorption coefficient, $\delta$ is the thickness of the absorber layer. The water can be absorbed from the ambient atmosphere onto the surface of the microresonator. Silicon is typically coated with a thin film of silicon dioxide, which is highly hydrophilic. It will influence optical losses, especially at wavelengths where the absorption in water is significant.

\section{Experimental setup}

\begin{figure}[htbp]
  \centering
  \includegraphics[width=1\linewidth]{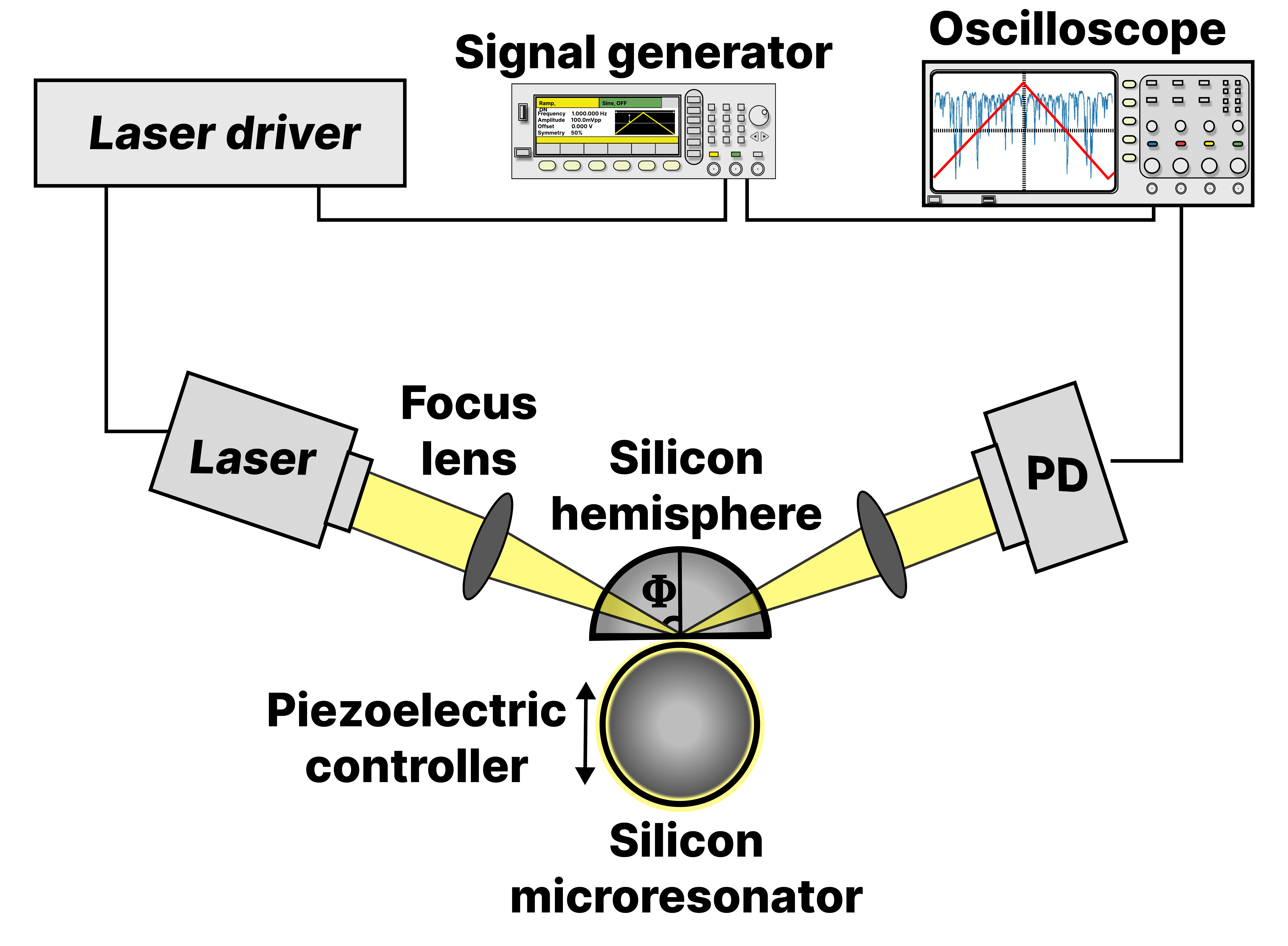}

\caption{Fig. 1. Experimental setup for Q-factor measurements. The pump wavelength was determined by the single-frequency laser used: fiber laser with isolator and polarization controller at 1.5 \textmu m, distributive feedback diode laser without isolator at 2.6 \textmu m, quantum-cascade laser without isolator at 6.1 \textmu m and 8.6 \textmu m), PD is the photodetector for the operating wavelength, $\Phi$ is the angle of incidence.}
  \label{fig:exp_setup}
\end{figure}

The experimental setup for Q-factor measurements is presented in Fig.~\ref{fig:exp_setup}. The pump wavelength depends on the used single-frequency laser (fiber laser with isolator and polarization controller at 1.5 \textmu m (Koheras Adjustik), diode laser with distributive feedback (DFB) without an isolator at 2.6 \textmu m (Nanoplus), quantum-cascade lasers (QCL) without isolators at 6.1 \textmu m (QD6500CM1) and 8.6 \textmu m (QD8650CM1)). The light was coupled into the microresonator through the focusing lens and coupling element. The three-coordinate translation stage with PZT was used to control the alignment, precise positioning, and adjustment of the coupling coefficient of the microresonator. The transmitted light was collected on the photodetector for the operating wavelength (DET10N/M by Thorlabs for 1.5 \textmu m, PD27 by IoffeLed for 2.6 \textmu m, PD65 by IoffeLed for 6.1 \textmu m and 8.6 \textmu m).

We excited WGM in silicon microresonators by means of a hemispherical coupler also made of silicon \cite{Shitikov:18}. It is a universal method for WGM excitation valid for all wavelengths discussed in this research. The angle of incidence ($\Phi$) of the pump light can be calculated using the following formula \cite{Shitikov:18}:

\begin{equation}
    \Phi\cong \frac{\pi}{2}-\sqrt{-\zeta_q}\left(\frac{m+p}{2}\right)^{-\frac{1}{3}}, 
\label{Gammam}
\end{equation}
where $\zeta_{q}$ is a root of the Airy function ($\zeta_{1,2,3,...}\simeq-2.338,-4.088,-5.521,...$), $q$ and $p$ are the radial and vertical mode indices respectively, $m\approx 2\pi a n/\lambda$ is the azimuthal index of the mode and $\lambda$ is the laser wavelength. The angle of incidence depends on $\lambda$ in terms of \textit{m}($\lambda$, \textit{n}($\lambda$)). The refractive index becomes insignificantly larger with increase in wavelength. On the other hand, $m$ inversely depends on $\lambda$. It affects the optimal angle of incidence (Tab. \ref{tab:angles}).

 \begin{table}[htbp]
\centering

\caption{\bf Tab. 1. Dependence of the optimal incident angle on the pump wavelength.}
\begin{tabular}{cccc}
\hline
$\lambda$, \textmu m &  \textit{n} \cite {Li:80} &  \textit{m} & $\Phi$, $^o$\\
\hline
1.5 & \,3.4799\, & \,18200\, & 85.8\\
2.6 & 3.4359 & 10400 & 84.9\\
6.1 &  3.4172 & 4200 & 83.3\\
8.6 & 3.4155 & 3100 & 82.5\\

\hline
\end{tabular}
  \label{tab:angles}
\end{table}

The frequency scan is made by triangular sweep of the laser diode current. It is calibrated by a free-space interferometer with a free spectral range (FSR) of 240 MHz at 1.5 \textmu m and germanium Fabry-Pérot  interferometers with FSR of 725 MHz at 2.6 \textmu m and 1.47 GHz for wavelengths of 6.1 and 8.6 \textmu m.

\section {Methods}

In this work, we have focused on measuring the quality factor and its dependence on wavelength. The pump power and conductivity provide sufficient information to draw conclusions about the loss mechanisms. Whispering gallery modes are localized in a vicinity of the circumference of the microresonator, thus they are sensitive to surface scattering determined by its roughness and absorption due to contamination as well as to the bulk scattering and absorption. Therefore, a common surface treatment technique should be implemented.

Identical microresonators are fabricated from different types of crystalline silicon. Resonators are disks with a thickness of approximately 1.5 mm, a diameter of 2.5 mm, and a radius of curvature of $(1\pm0.2)$ mm. Then, all the microresonators were manually polished. Asymptotic polishing with diamond slurries was used for the surface treatment of the resonators \cite{Ilchenko_nonlinear_2004}. The final polishing was performed with OXAPA\textregistered \, colloidal silica slurry with a particle size less than 50 nm. The polishing technique is crucial for reducing surface losses. It was demonstrated that the Q-factor exceeding $10^9$ at 1.5 \textmu m can be achieved in crystalline silicon using polishing with silicon dioxide colloidal slurry \cite{Shitikov:18} which corresponds to optical losses below 100 ppm/cm. The uniform method for preparing microresonators allows us to make a comprehensive analysis of the reasons affecting the optical losses in silicon microresonators.

\begin{figure*}[htbp!]
  \centering
  \includegraphics[width=0.6\linewidth]{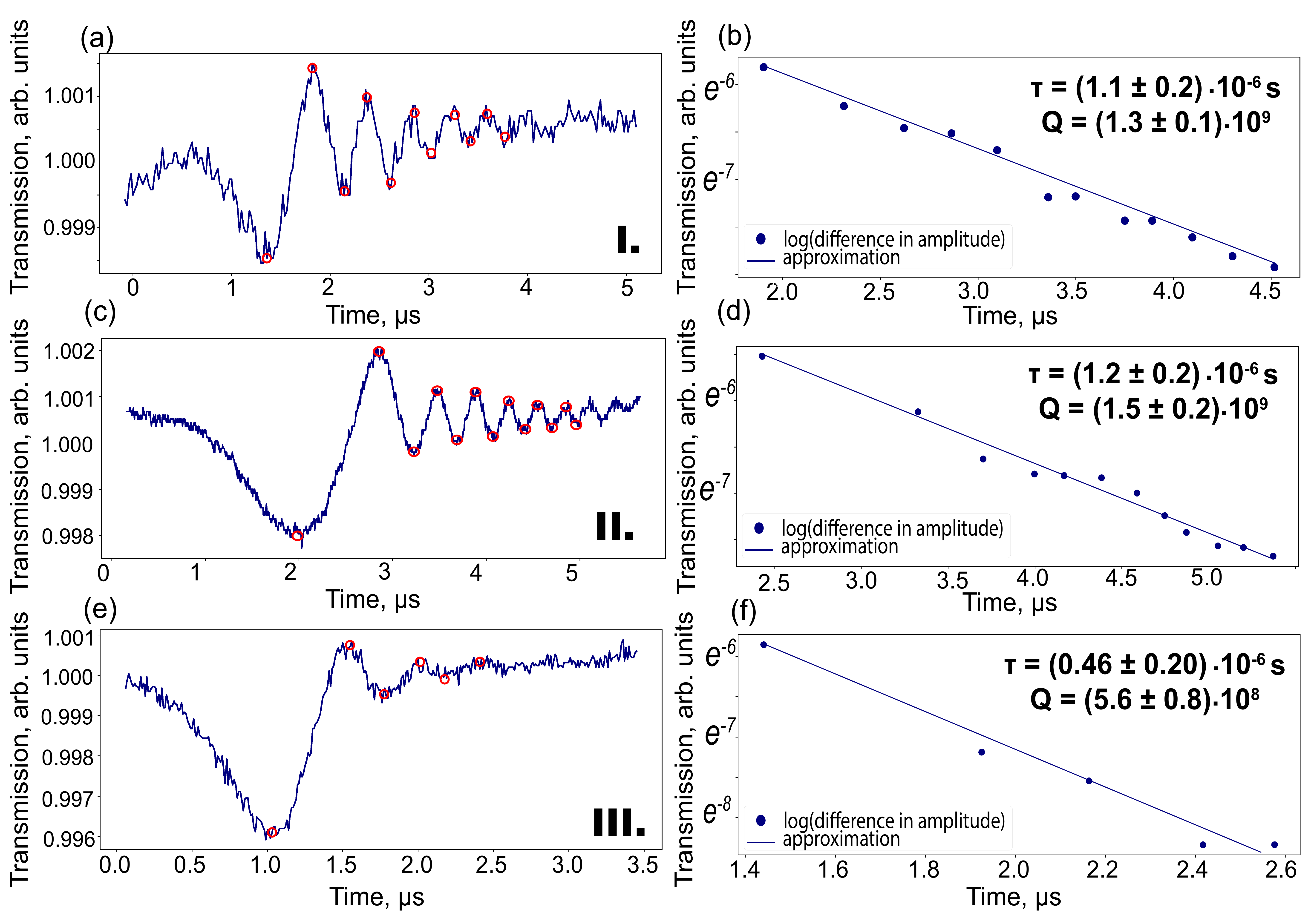}

\caption{Fig. 2. Ring-down measurements of samples 1-3 at 1.5 $\mu$m. In the left column transmission signals are presented. In the right column the exponential approximations are presented. The measurements are conducted in the undercoupled regime. }
  \label{fig:RD}
\end{figure*}

We used three different methods for precise quality factor measurements. The first method is based on the recording of a resonance curve with subsequent fitting with a Lorentzian profile and determination of the full width at half maximum (FWHM) of the resonance curve. It gives excellent results if the linewidth of the pump laser and the fluctuations of the laser frequency during the measurement are much smaller than the width of the resonance curve.
Resonances are recorded twice, during forward and backward frequency scans. The coincidence of the two FWHM verifies the absence of nonlinear effects. 
The measurements are performed in either a critical coupling regime (where the coupling losses are equal to the internal losses of the microresonator) or a weak coupling regime (where the coupling losses are negligible).
For extremely high Q-factor resonances, laser frequency noise perturbs the resonance curve, or the nonlinearity may become significant. This is usually evident in different FWHM of the resonance curve in forward and backward scans. In this case, a method based on the ring-down technique is used. The laser frequency is rapidly tuned across the WGM eigenfrequency. As a result, decaying oscillations due to interference between damped oscillations in the WGM and the detuned laser pump are observed in the transmission spectrum. Exponential fitting allows us to evaluate the decay rate and estimate the Q-factor. A drawback of this method is the need for a high-speed photodetector. Such a detector with 400 MHz bandwidth is used for 1.5 \textmu m measurement, but it was not available for a longer wavelength. 
For measuring high-Q factor at 2.6 \textmu m, we use a method based on the self-injection-locking (SIL) effect. Measurements of the dependence of the locking range on the gap between the microresonator and the coupling element allows us to evaluate Q-factor as well as the excited mode's angular order \cite{shitikov2020microresonator}.
The advantages of this method are its low sensitivity to laser noise and the ability to use slow photodetectors. The disadvantage is the need to measure the gap between the microresonator and the coupling element with nanometer accuracy.

At long wavelengths, optical losses may increase with a possible contamination by dust particles during the experiment. To ensure that scattering losses remain consistent and that the Q-factor does not degrade, we verify measurements at 1.5 \textmu m after each series of measurements of the Q-factor in the mid-IR band.

\begin{figure*}[htbp]
  \centering
  \includegraphics[width=\linewidth]{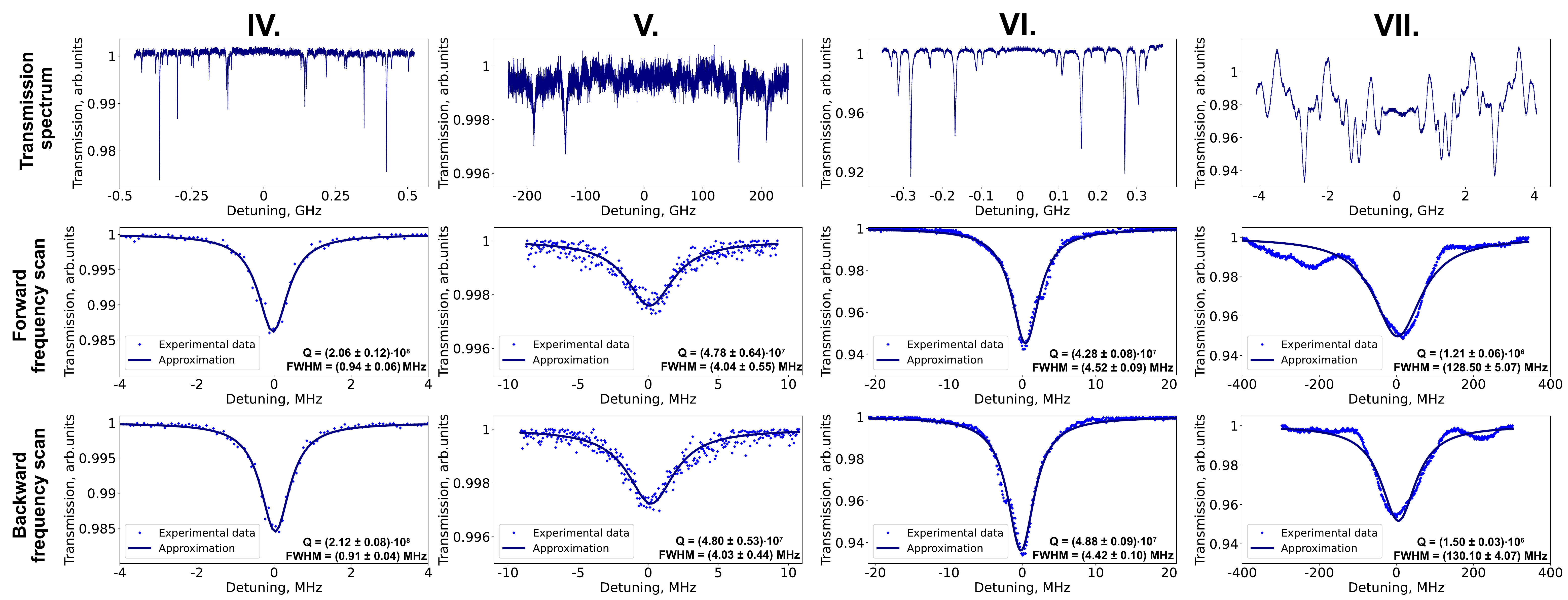}

\caption{Fig. 3. The full-width at half maximum measurements of the samples 4-7 at 1.5 $\mu$m. In the first row the transmission spectra of the forward and backward frequency scans are presented. In the middle and bottom row the Lorentzian approximations of the resonances are presented.  }
  \label{fig:1550}
\end{figure*}

\section{Q-factor measurements}

Several types of silicon samples with different resistivity are tested; see Tab. \eqref{tab:shape-functions}. Samples 1-3, 6, and 7 are grown by the high-floating zone (HFZ) technique. Samples 4 and 5 are grown using the magnetic Czochralsky (MCZ) technique \cite{hoshi1985czochralski}.

{\bf  1.5 \textmu m measurements}

First, the Q-factor for seven silicon microresonators is measured at 1.5 \textmu m (see Tab. \ref{tab:shape-functions} and Figs. \ref{fig:RD} and \ref{fig:1550}). The first column in Tab. \ref{tab:shape-functions} is a sample number, the second for $\rho$ is the resistivity measured by the supplier, the next is a crystal growth method and conductivity type, $\alpha_{int}$ is the optical losses measured in \cite{Losses}, $\alpha_{exp}$ is the optical losses evaluated from the Q, the last column is the values of the measured Q-factor.   

\begin{table}[htbp]
\centering

\caption{\bf Tab. 2. Comparison of measured optical losses for crystalline silicon microresonators with different resistivity measured at 1.5 \textmu m. $^*$NA - no available information}
\begin{tabular}{ccccc}
\hline
N & $\rho$,& Method,& $\alpha_{int} $, & Q\\
 & $\rm{k\Omega \cdot m}$&  type  & $\rm{ppm/cm}$ \cite{Losses} & \\
\hline
$1$ & 8 & HFZ, n & 2.3 - 5  & 1.3$\cdot10^9$\\
$2$& 37 & HFZ, p  & 2.8 - 9.2 &  1.5$\cdot10^9$\\
$3$ & 2.2 & HFZ, n & 37  & 5.6$\cdot10^8$\\
$4$ & 6 & MCZ, NA$^*$ & 4-6 & 2$\cdot10^8$\\
$5$ & 4.4& MCZ, NA$^*$  & 4-6 &  5$\cdot10^7$\\
$6$ & 0.1 & HFZ, p  & 775 & 4$\cdot10^7$\\
$7$ & 0.09& HFZ, p & 803 & 1.5$\cdot10^6$\\

\hline

\end{tabular}
  \label{tab:shape-functions}
\end{table}

The 1st and 3rd samples have n-type conductivity and the 2nd, 6th and 7th samples have p-type conductivity. For the 4th and 5th samples there is no information about conductivity type. Thus, the 1st - 3rd samples have almost the same Q-factor, and we conclude that the conductivity type does not noticeably affect the optical losses. At the same time, it is worth noting, that almost equal resistivity the Q-factors of the MCZ samples are 10 times lower than for the HFZ samples. 

\begin{figure}[htbp!]
  \centering
  \includegraphics[width=0.75\linewidth]{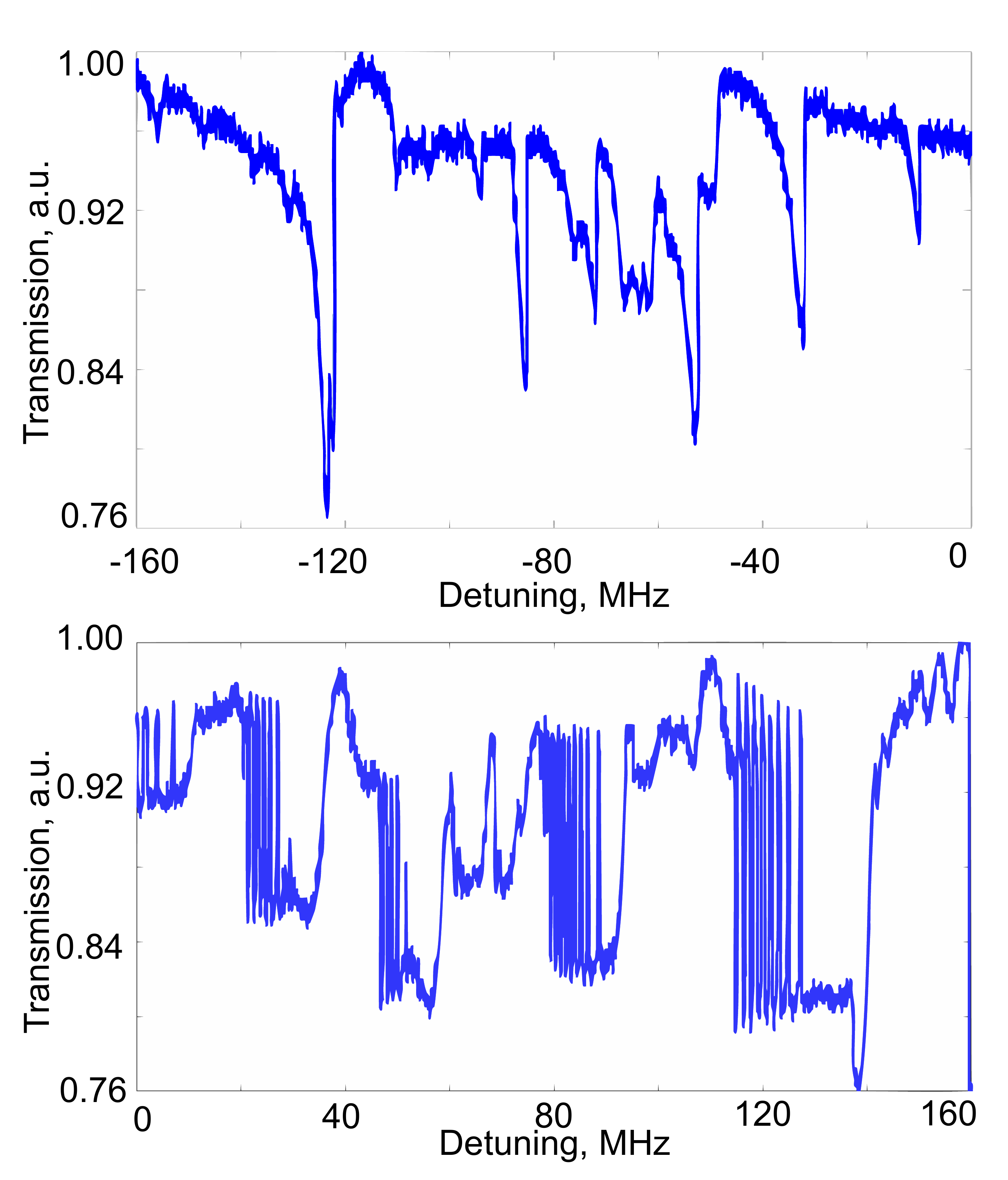}

\caption{Fig. 4. The transmission spectrum for high-power pumping of the 3d sample in the critical coupling regime. The distinct thermo-optic oscillations are observed.}
  \label{fig:HP}
\end{figure}

In Fig. \ref{fig:HP} the transmission spectrum is presented in the case of high power pumping ($\approx$100 mW) of the 3d sample in the critical coupling regime. The transmission spectrum is presented for the increase and decrease of the pump frequency. The resonance curves become highly nonlinear and strong thermo-optic oscillations are observed.

\begin{figure}[htbp!]
  \centering
  \includegraphics[width=\linewidth]{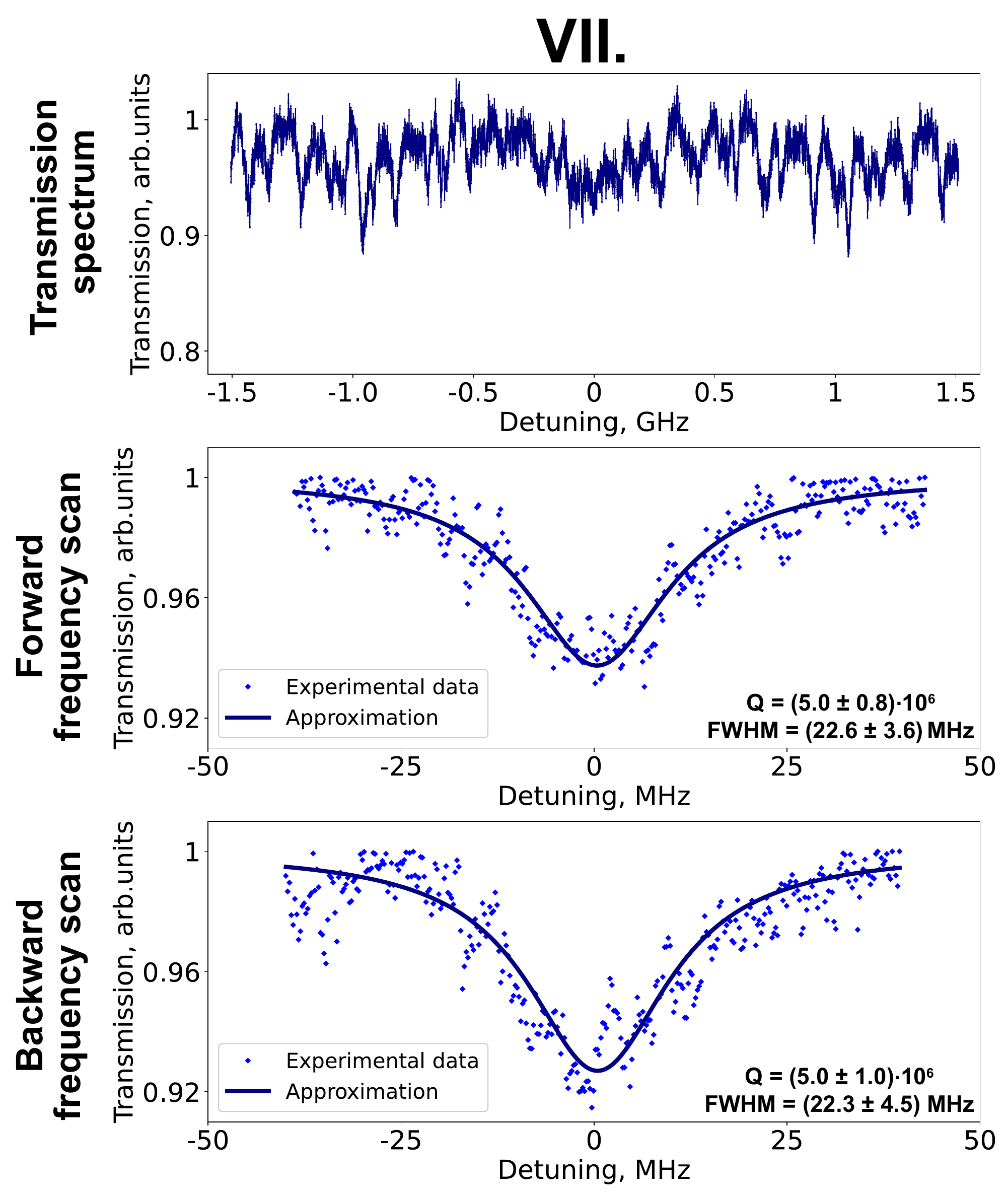}

\caption{Fig. 5. The full-width at half maximum measurements for the 7th microresonator at 2.6 $\mu$m. In the first row the transmission spectra of the forward and backward frequency scans are presented. In the middle and bottom row the Lorentzian approximations of the resonances are presented.}
  \label{fig:2_6}
\end{figure}

{\bf  2.6 \textmu m measurements}

To measure the Q factor at 2.6 \textmu m, we use the FWHM-based method for the 6th and 7th microresonator and the self-injection locking (SIL) based method for the 3d and 4th microresonator since the FHWM appears to be less than the linewidth of the laser. The measured Q-factor of the 3d sample was 2$\cdot10^8$. That is lower than expected taking into account Q = 5$\cdot10^8$ in the telecommunication band. However, most of the resonances at 1550 nm had $Q\approx10^8$. Such Q-factor is reasonable for them. 
In our previous study \cite{shitikov2020microresonator}, we have already measured the quality factor of the 4th microresonator which was $5\cdot10^8$ (see Tab. \ref{tab:all_wl}). Here, in this work, we verified this value. 
The intrinsic Q-factor for the 6th sample was 4.4$\cdot10^7$ almost the same as at the telecommunication wavelength. The measured loaded Q-factor for the 7th microresonator in the critical coupling regime is $5\cdot10^6$, therefore, the intrinsic Q-factor is $Q = 1\cdot10^7$. The transmission spectrum for the 2.6 \textmu m measurements is presented in Fig. \ref{fig:2_6}. 

\begin{figure*}[htbp!]
  \centering
  \includegraphics[width=0.65\linewidth]{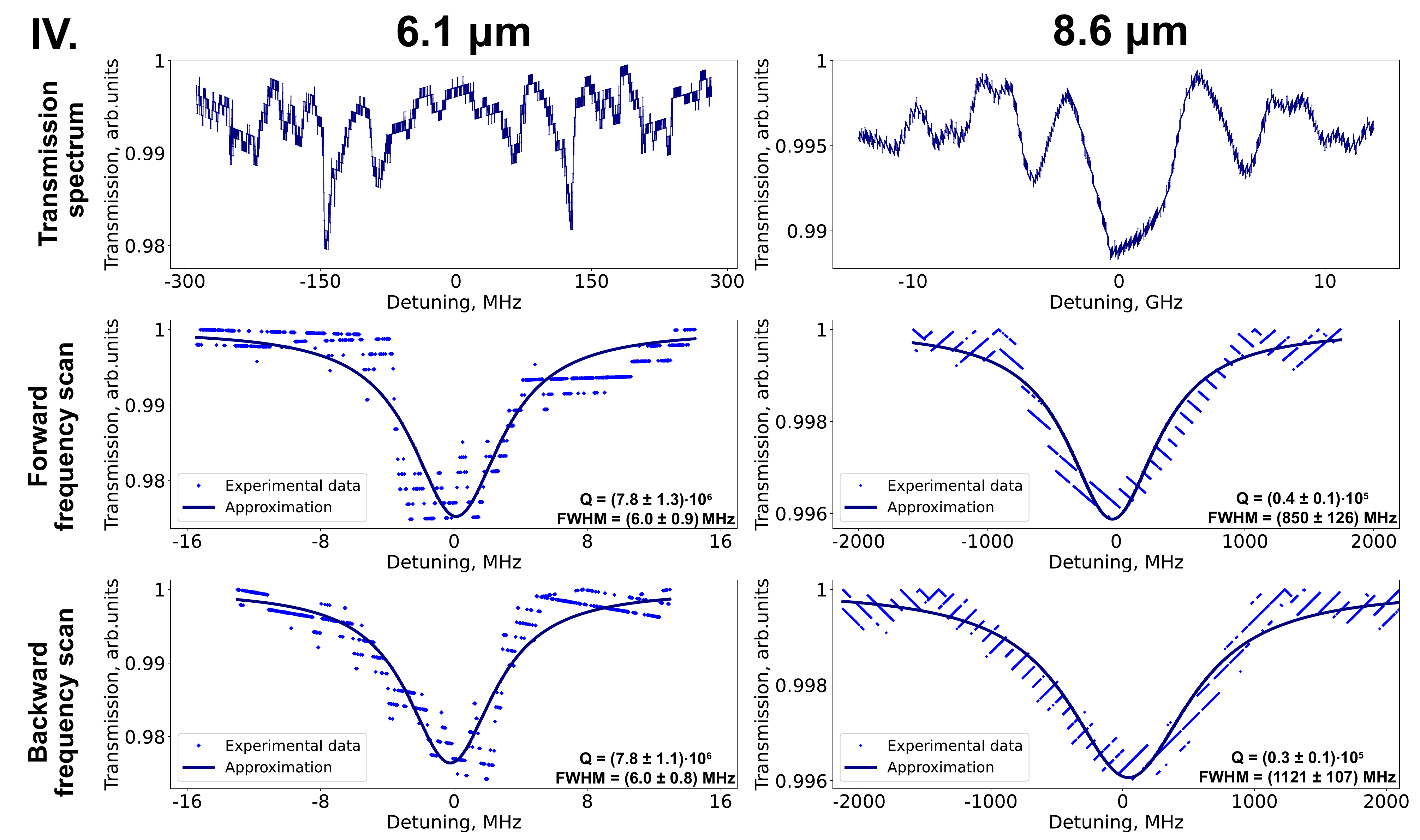}

\caption{Fig. 6. The full-width at half maximum measurements for the 4th microresonator measured at 6.1 and 8.6 $\mu$m. In the first row the transmission spectra of the forward and backward frequency scans are presented. In the middle and bottom row the Lorentzian approximations of the resonances are presented.  }
  \label{fig:6_8}
\end{figure*}

It should be emphasized that, despite the difference in conductivity type, resistance and growth method, resonators reproducibly demonstrate a quality factor at 2.6 \textmu m higher than at 1.5 \textmu m. It is associated with the decrease of Rayleigh scattering.

{\bf  6.1 \textmu m measurements}

For Q-factor measurements at 6.1 \textmu m wavelength we use FWHM-based method. 
The intrinsic Q for the 4th microresonator measured at 6.1 \textmu m is $1.6\cdot 10^7$ and for the 7th is $4\cdot 10^6$.
The transmission spectra are represented in Fig. \ref{fig:6_8}.
It is worth noting that the Q-factor was also determined for the 3d sample. It reached 4$\cdot10^6$ but the resonance curve demonstrated the signs of a SIL regime, namely sharp edges and a cyclical change in the resonance curve with the distance between the laser and the microresonator\cite{shitikov2023optimization}. So that, SIL might lead to the broadening of the resonance curve. However, the SIL regime is mostly the parasitic effect for microresonator Q-factor measurements. Comparing to the successful realization of SIL in mid-IR \cite{doi:10.1002/lpor.201500214, Savchenkov:15}, here, the wavelength is longer and the coupling element has large reflection coefficient, so we did not see stable SIL regime. 

{\bf  8.6 \textmu m measurements}

We provide Q-factor measurements with the use of the FWHM-based method.
At 8.6 \textmu m an attempt is made to excite WGM in each of the microresonators from Tab. \ref{tab:shape-functions}, but only one successful result is obtained with the 4th microresonator made of silicon grown by the Czochralsky technique with resistivity of 6 k$\Omega\cdot$ m. The loaded Q-factor in this case is 4$\cdot 10^4$, intrinsic is $Q = 8\cdot 10^4$ (see Fig. \ref{fig:6_8}). This dramatic Q-factor degradation apparently connected to multi-phonon absorption.

{\bf  Comparison}

For the comparison, we choose the 4th and 7th microresonators from the Tab. \ref{tab:shape-functions}. The choice is motivated by the desire to compare microresonators made of silicon with different resistivity and optical losses. 

Results obtained across all the wavelengths are presented in Tab. \ref{tab:all_wl}. 
The Q-factor is maximal at 2.6 \textmu m for both microresonators, suggesting that dopant-induced absorption has a minimum at this wavelength. 
\begin{table}[htbp]
\centering

\caption{\bf Tab. 3. Comparison of measured Q-factor for crystalline silicon microresonators with different resistivity}
\begin{tabular}{ccccc}
\hline
N & 1.5 \textmu m & 2.6 \textmu m & 6.1 \textmu m & 8.6 \textmu m\\
\hline

$4$ & 2$\cdot10^8$ & $5\cdot10^8$  & 1.6$\cdot10^7$ & 0.8$\cdot10^5$\\
$7$ & 1.5$\cdot10^6$ & 1$\cdot10^7$ & 4$\cdot10^6$& - \\

\hline
\end{tabular}
  \label{tab:all_wl}
\end{table}

The Q-factor decreases from its maximum at 2.6 \textmu m to its lowest values at 6.1 and 8.6 \textmu m. According to Eq. \ref{eq:alphass} the scattering is suppose to decrease as $\lambda^4$. The multiphoton absorption and FCA also suppose to decrease \cite{wang2013multi}. The increase of the optical losses may be affiliated with the water absorption peak at 6.1 \textmu m compared to the absorption at shorter wavelengths previously discussed in this study (10.834 cm$^{-1}$ at 1.5 \textmu m, 122.00 cm$^{-1}$ at 2.6 \textmu m, 2660 cm$^{-1}$ at 6.1 \textmu m, 543.46 cm$^{-1}$ at 8.6 \textmu m) \cite{segelstein1981complex}. Using Eq. \ref{eq:alphawa} we estimate the thickness of the water layer for Q=$1.5\cdot10^7$ as $2\cdot10^{-12}\,\rm{m}$ at 6.1 \textmu m. On the other hand, the thickness of water layer that would limit the Q-factor $1.5\cdot10^9$ at 1.55 is $3\cdot10^{-12}\,\rm{m}$. So that, the value is proportional and the water absorption can be the main reason of the Q-factor limitations. The water absorption coefficient for 8.6 \textmu m is also significant, but it cannot be the main reason of the high optical losses.

We take into account the possible reason of the Q-factor reduction at longer wavelengths connected to the increase in the mode volume, which may become large enough to include the unpolished areas. The polished belt width which is approximately 100 \textmu m, is controlled with a microscope to ensure that even at 8.6 \textmu m surface roughness does not affect the Q-factor. 
The most likely mechanism for additional losses is a multi-phonon absorption peak at 8.6 \textmu m. 

At wavelengths of 1.5 \textmu m, 2.6 \textmu m, and 6.1 \textmu m, the low-resistivity microresonators exhibit lower Q-factors than the high-resistivity ones.

\section{Conclusion}

In summary, we measure the Q-factor of microresonators made of different types of crystalline silicon at wavelengths of 1.5 \textmu m, 2.6 \textmu m, 6.1 \textmu m and 8.6 \textmu m. To the best of our knowledge, the Q-factors measured in the high-resistivity resonators are record values. We obtain 1.5$\cdot 10^9$ at 1.5 \textmu m, 5$\cdot 10^8$ at 2.6 \textmu m, 1.6 $\cdot 10^7$ at 6.1 \textmu m, and 8$\cdot10^4$ at 8.6 \textmu m. 
We show that the optical losses have a tendency to decrease in the range of wavelengths from 1.5 to 2.6 \textmu m, then increase for wavelengths 6.1 \textmu m and 8.6 \textmu m. We explain the decrease of the optical losses from 1.5 to 2.6 \textmu m with the decrease of Rayleigh scattering. Then, we suppose, the degradation of Q-factor at 6.1 \textmu m might be connected with water absorption. 
Finally, the optical losses at 8.6 \textmu m are mainly determined by multi-phonon absorption. We manage to excite WGM only in one type of the microresonators and measure the Q-factor of this microresonator as $8 \cdot 10^4$.

We show that the conductivity type does not noticeably influence the optical losses while the resistivity is a defining factor. Samples made by the high-floating zone technique demonstrate a tenfold Q-factor compared to those made by the Czochralsky technique with the same resistivity.
Our study confirms the utility of WGM microresonators as loss measurement tools in wide range of wavelength and provides significant potential for the development of silicon microresonator-based photonics in the mid-IR band.

\section{Acknowledgment}
The work was supported by Russian Science Foundation (project 25-12-00263).
The experimental characterization of samples 1 and 2 at 1550 nm was supported by Rosatom in the framework of the Roadmap for Quantum computing (Contract No. 868-1.3-15/15-2021 dated October 5).

\eject
\bibliography{Textbib}

\end{document}